\documentclass[11pt]{article}
\usepackage{moriond,epsfig}
\usepackage{graphicx}
\def\Journal#1#2#3#4{{#1} {\bf #2}, #3 (#4)}

\def\NIMA{{\em Nucl. Instrum. Methods} A}

\def\PLB{{\em Phys. Lett.}  B}
\def\PRL{\em Phys. Rev. Lett.}

\def\NJP{\em New J.Phys}
\def\TNS{\em IEEE Trans. Nucl. Sci.}
\def\mco{\multicolumn}

\def\be{\begin{equation}}
\def\ee{\end{equation}}
\def\bea{\begin{eqnarray}}
\def\eea{\end{eqnarray}}

\begin{document}
\title{\Large Observation of neutrino interactions in the OPERA detector }
\author{{\bf Alberto Garfagnini} \\ {\em Dipartimento di Fisica dell'Universit\`a di Padova and INFN, \\I-35131 Padova, Italy} }

\author{{\bf Ciro Pistillo} \\ {\em Universit\"{a}t Bern, Laboratorium f\"{u}r Hochenergie Physik,\\ Sidlerstrasse 5, CH-3012 Bern, Switzerland \\ \hspace {1cm} \\ \hspace{1cm}}}

\maketitle\abstracts{OPERA is a long baseline neutrino oscillation experiment designed to observe $\nu_{\mu} \rightarrow \nu_{\tau}$ oscillations by looking at the appearance of $\nu_{\tau}$'s in an almost pure $\nu_\mu$ beam. The beam is produced at CERN and sent towards the Gran Sasso INFN laboratories where the experiment is running. OPERA started its data taking in October 2007, when the first 38 neutrino interactions where successfully located and reconstructed.
This paper reviews the status of the experiment discussing its physics potential and performances for neutrino oscillation studies.}
\vspace{5mm}
\section{Introduction}
OPERA~\cite{opera_proposal} is a long baseline experiment
at the Gran Sasso underground laboratories (LNGS) and is part
of the CERN Neutrino to Gran Sasso (CNGS) project.
The detector has been designed to observe the
$\nu_{\mu} \rightarrow \nu_{\tau}$ oscillations in the parameter region
indicated by Super-Kamiokande~\cite{Fukuda:1998mi} through direct
observation of $\nu_{\tau}$ charged current interactions.
The detector is based on a massive lead/nuclear emulsion target
complemented by electronic detectors (scintillator bars) that allow
the location of the event and drive the scanning of the emulsions.
A magnetic spectrometer follows the instrumented target and measures
the charge and momentum of penetrating tracks.
%The target and the spectrometer constitutes one ``supermodule''.
%OPERA is made of two supermodules and is installed in Hall C of LNGS.

\section{The CNGS neutrino beam}
The Cern Neutrino to Gran Sasso(CNGS)~\cite{cngs}
facility is a wide-band neutrino beam which provides an almost pure $\nu_\mu$
source traveling 730 km under the Earth crust from CERN to Gran Sasso.
The beam parameters have been designed in order to optimize the
number of $\nu_\tau$ charge current interactions in the OPERA detector.
The neutrino beam mean energy is $<E_\nu> = 17~\mbox{GeV}$
with a very small $\nu_e$ and $\overline{\nu_e}$ contamination (less than 1\%).
The average $L/E$ ratio is 43 km/GeV, far from the oscillation maximum,
but dictated by the high energy needed for $\nu_\tau$ appearance.
The beam has been designed to provide $45 \cdot 10^{18}$ proton-on-target/year
(p.o.t./y) with a running time of 200 days per year.

The first CNGS technical run occurred in August 2006 with a delivered
luminosity of $0.76 \cdot 10^{18}$ p.o.t.
A new short run followed in October 2006, but was shortly interrupted due to
a leak in the closed water cooling system of the reflector: only
$0.06 \cdot 10^{18}$ p.o.t. were delivered for the experiment.
At that time only the electronic detectors were installed and under
commissioning.

After repair of the reflector cooling system, a new physics run
occurred in October 2007, when OPERA had 40\% of the target mass installed.
The beam extraction intensity was limited to 70\% of the normal values
due to beam losses which brought severe radiation damage of the
equipment. Due to these new technical problems,
only $0.79 \cdot 10^{18}$ p.o.t. were delivered. The beam operation was
interrupted due to loss of ventilation control in the CNGS area due to the
radiation damage of the CNGS standard electronics.

A major revision of the project has been taken in the beginning of 2008
in order to improve the radiation shielding of the electronics and
reduce the beam losses. A new physics run is going to start in summer
2008 with a planned luminosity of $~\sim 30 \cdot 10^{18}$ p.o.t.
for the CNGS experiments in Gran Sasso.

\section{The OPERA detector}
\begin{center}
\begin{figure}[ht]
  \hspace{2.5cm}
  \includegraphics[scale=0.8]{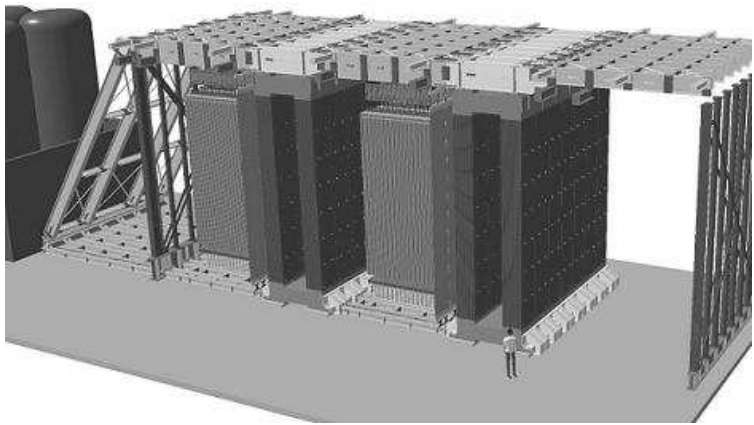}
  %\special{psfile=LaThuileFPSprofig.ps voffset=-220 hoffset=-30
  %  hscale=80 vscale=80 angle=0}
  \caption{\it Schematic view of the OPERA detector. The neutrino beam enters
               the detector from the left.}
    \label{fig:opera_draw}
\end{figure}
\end{center}

OPERA is a large detector ($10~\mbox{m} \times 10~\mbox{m} \times 20~\mbox{m}$)
located in the underground experimental Hall C of LNGS.
As shown in Figure~\ref{fig:opera_draw},
the detector is made of two identical super-modules, aligned along the CNGS beam
direction, each one consisting of a target and a muon spectrometer.
The target section combines passive elements, the lead-emulsion bricks, and
electronic detectors. Each target section consists of a multi-layer
array of 31 target walls followed by pairs of planes of plastic
scintillator strips (Target Tracker).
A magnetic spectrometer follows the instrumented target and measures
charge and momentum of the penetrating tracks.

\subsection{The Emulsion Target}
The development of automatized scanning systems during the last two decades has made 
possible the use of large nuclear emulsion detectors.
Indeed, nuclear emulsion are 
still successfully used nowadays, especially in neutrino
experiments~\cite{Chorus}~\cite{Donut}.
The realization of a new scanning system has been carried out by two different R\&D programs in the Nagoya University (Japan) and in several european laboratories belonging to the OPERA collaboration. These scanning systems~\cite{SUTS}~\cite{ESS2}~\cite{ESS1} were designed to take into account the requests of high scanning speed (about 20 cm$^2$/h) while keeping the extremely good accuracy provided by nuclear emulsions. About 40 automatic microscopes are installed in the various scanning laboratories of the OPERA experiment. 

The total number of emulsion films in the OPERA detector will be about 9 millions, for an area of about 110000 m$^2$. These quantities are orders of magnitude larger than the ones used by previous experiments. That made necessary an industrial production of the emulsion films, performed by the Fuji Film company, in Japan, after an R\&D program conducted jointly with the OPERA group of the Nagoya University.

The OPERA emulsions are made up of two emulsion layers 44 $\mu$m thick coated on both sides of a 205 $\mu$m triacetate base. The AgBr crystal diameter is rather uniform, around 0.2 $\mu$m, and the sensitivity is about 35 grains/100 $\mu$m for minimum ionizing particles.

The main constituent of the OPERA target is the brick. It is a pile of 57 nuclear emulsion sheets interleaved by 1 mm thick lead plates. The brick combines the high precision tracking capabilities provided by the emulsions with the large mass given by the lead. The OPERA brick is a detector itself. In addition to the vertex identification and $\tau$ decay detection, shower reconstruction and momentum measurements using the Multiple Coulomb Scattering can be performed, being the total brick thickness of 7.6 cm equivalent to 10 X$_0$. Bricks are hosted in the walls of the target. 

The occurrence of a neutrino interaction inside the target is triggered by the electronic detectors. Muons are reconstructed in the spectrometers and all the charged particles in the target tracker. The brick finding algorithm indicates the brick where the interaction is supposed to be occurred. The trigger is confirmed in the Changeable Sheet Doublet (CSD)~\cite{bib:csd_paper}, a pair of emulsion films hosted in a box placed outside the brick, as interface between the latter and the target tracker. Before detaching the CSD from the brick, they are exposed to an XRay spot, in order to define a common reference system for the two CS and the first emulsion in the brick (with a precision of a few tens of $\mu$m). Afterwards the CS are developed and the predictions from target tracker are searched for within a few cm area. If these are confirmed the brick is brought outside the Gran Sasso laboratory and exposed to cosmic ray before development. \\
The mechanical accuracy obtained during the brick piling is in the range of 50-100 $\mu$m. The reconstruction of cosmic rays passing through the whole brick allows to improve the definition of a global reference frame, leading the precision to about 1-2 $\mu$m. \\
All the tracks located in the CSD are subsequently followed inside the brick, starting from the most downstream film, until they stop. Then a general scanning around the stopping point(s) is performed, tracks and vertices are reconstructed, the primary vertex is located and the kinematic analysis defines the event topology.

\subsection{The Target Tracker}
The main role of the Target Tracker is to provide a trigger
and identify the right bricks where the event
vertex should be located.
Each wall is composed by orthogonal planes of plastic scintillator
strips ($680~\mbox{cm} \times 2.6~\mbox{cm} \times 1~\mbox{cm}$).
The strips are made of extruded polystyrene with 2\% p-terphenyl and
0.02\% POPOP, coated with a thin diffusing white layer of TiO$_2$.
Charged particle crossing the strips will create a blue scintillation light
which is collected by wavelength-shifting fibers which propagate light
at both extremities of the strip. All fibers are connected at both ends
to multianode Hamamatsu PMTs.
For a minimum ionizing particle, at least five photoelectrons
are detected by the photomultipliers.
The detection efficiency of each plane is at 99\%.
A detailed description of the Target Tracker design can be found
in~\cite{adam:2007ex}.

\subsection{The Spectrometer}
The spectrometer allows to suppress the
background coming from charm production through the identification of
wrong-charged muons and contributes to the kinematic reconstruction of the
event performed in the target section.
\begin{figure}
\centering
\includegraphics[width=3.5in]{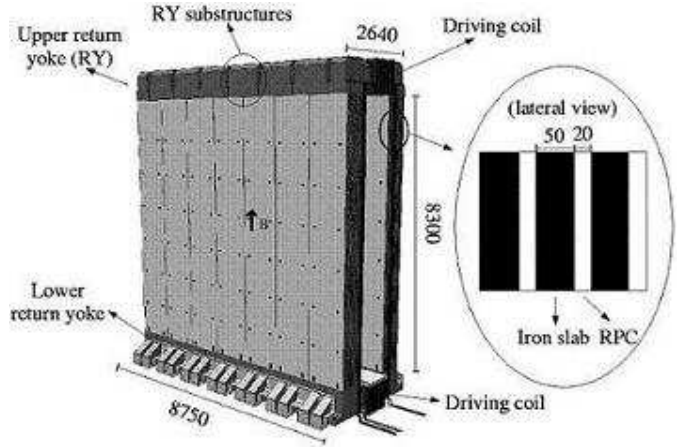}
\caption{\it {Three dimensional view of the OPERA magnet. Units are in mm.}}
\label{fig:magnet}
\end{figure}
The magnet~\cite{Ambrosio:2004tm},
shown in Figure~\ref{fig:magnet}, is made of two vertical
walls of rectangular cross section and of a top and bottom flux
return path.
The walls  are built lining twelve iron layers (5 cm thickness)
interleaved with 2 cm of air gap, allocated for the housing of
the Inner Tracker detectors, Resistive Plate Chambers, RPCs.
Each iron layer is made of seven slabs, with dimensions
$50 \times 1250 \times 8200~\mbox{mm}^3$, precisely milled along
the two $1250~\mbox{mm}$ long sides connected to the return yokes
to minimize the air gaps along the magnetic circuit.
The slabs are bolted together to increase the compactness and the
mechanical stability of the magnet which acts as a base for the
emulsion target support.
The nuts holding the bolts serve as spacers between two slabs and
fix the 20~mm air gap where the RPCs are mounted.

%\begin{center}
\begin{figure}[ht]
  % \vspace{9.0cm}
  \begin{center}
  \includegraphics[scale=0.75]{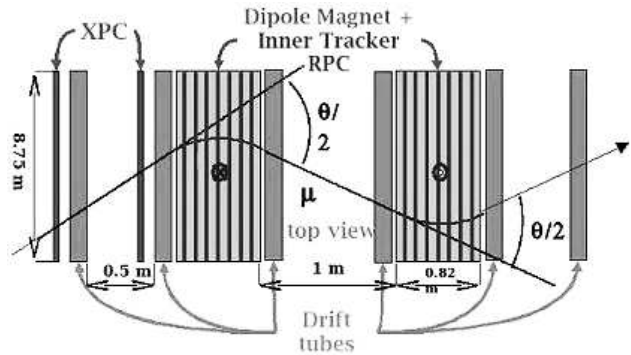}
  \end{center}
  %\special{psfile=LaThuileFPSprofig.ps voffset=-220 hoffset=-30
  %  hscale=80 vscale=80 angle=0}
  \caption{\it Top view of the OPERA muon spectrometer.
               The picture shows a track trajectory along the drift
               tube chambers, the XPCs and the RPCs inside the magnet
               ($dE/dx$ losses are neglected).}
    \label{fig:spectrometer_view}
\end{figure}
%\end{center}

The precision tracker is made of drift tubes planes located in front,
behind and between the two magnet walls: in total 12 drift tube planes
covering an area of $8~\mbox{m} \times 8~\mbox{m}$.
The tubes are 8 m long and have an outer diameter of 38 mm.
The trackers allows to reconstruct the muon momentum with a resolution
$\Delta p/p \leq 0.25$.

As shown in Fig.~\ref{fig:spectrometer_view}, a particle
entering the spectrometer is measured by layers of vertical drift tube
planes located before and after the magnet walls.
Left-right ambiguities are resolved by the two dimensional measurement
of the spectrometer RPCs and by two additional RPC planes, equipped
with pickup strips
inclined of
$\pm 42.6^\circ$ with respect to the horizon (XPC).
The Inner Tracker RPCs, eleven planes per spectrometer arm,
give a coarse measurement of the tracks and perform pattern recognition
and track matching between the precision trackers.
The OPERA RPCs~\cite{Bergnoli:2005cv}
are ``standard'' bakelite RPCs, similar to those used in the
LHC experiments: two electrodes,
made of 2 mm plastic laminate (HPL)
% with high volume resistivity
%($\rho \approx 10^{11} - 10^{12} \Omega$ cm),
are kept 2 mm apart by means of polycarbonate spacers in a 10 cm lattice
configuration.
The double coordinate readout is performed by means of copper strip panels.
The strip pitch is $3.5~\mbox{cm}$ for the horizontal strips and
$2.6~\mbox{cm}$ for the vertical layers.
The OPERA RPCs have a rectangular shape, covering an area of about 3.2 m$^2$.
The sensitive area between the iron slabs ($8.75 \times 8~\mbox{m}^2$),
is covered by twenty one RPCs arranged on seven rows, each with three RPCs
in a line.
In total, 1008 RPCs have been installed in the two spectrometers.

%\begin{center}
\begin{figure}[ht]
  \begin{center}
  \includegraphics[scale=0.25]{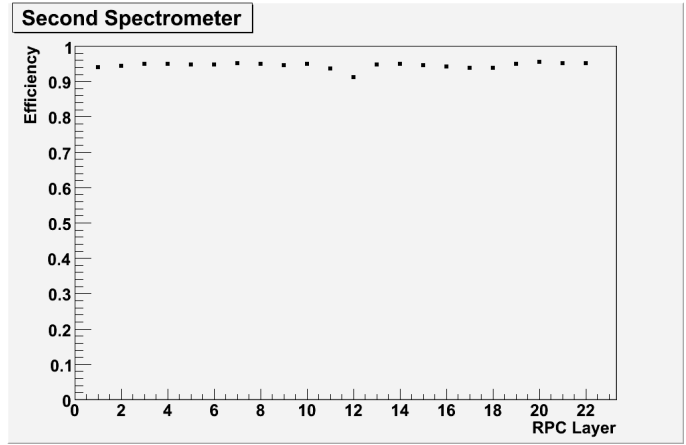}
  \end{center}
  \caption{\it Inner tracker plane efficiency. The mean value per plane,
               averaged over 21 RPCs, is shown.}
    \label{fig:rpceff}
\end{figure}
%\end{center}

Recent analysis of 2007 data, both with cosmic and beam events show
an average efficiency of 95\% for the RPC planes. Figure~\ref{fig:rpceff}
shows the average efficiency for the 22 layers of the second spectrometer~\cite{Dusini:opera_note86}.

\section{Physics performances}
The OPERA detector will host 155000 bricks for a total target mass of 1350 tons.
The signal of the occurrence of $\nu_{\mu} \rightarrow \nu_{\tau}$ oscillation is the charged current interaction of the $\nu_{\tau}$'s inside the detector target ($\nu_{\tau}N \rightarrow \tau^- X$). The reaction is identified by the detection of the $\tau$ lepton in the final state through the decay topology and its decay modes into an electron, a muon, and a single or three charged hadrons:

\begin{center}
\hspace{-1.84cm}$ \tau^- \rightarrow e^- \nu_{\tau} {\overline{\nu_e}} $ \\
\hspace{-1.74cm}$ \tau^- \rightarrow \mu^- \nu_{\tau} {\overline\nu_{\mu}} $\\
$ \tau^- \rightarrow (h^-h^+)h^-\nu_{\tau}(n\pi^0) $\\
\end{center}

The branching ratio for the electronic, muonic and hadronic channel are 17.8\%, 17.7\% and 64.7\% respectively. For the typical $\tau$ energies expected with the CNGS spectrum the average decay length is $\sim 450~\mu\mbox{m}$.

Neutrino interactions will occur predominantly inside lead plates. Once the $\tau$ lepton is produced, it will decay either within the same plate, or further downstream. In the first case, $\tau$ decays are detected by measuring the impact parameter of the daughter track with respect to the tracks originating from the primary vertex, while in the second case the kink angle between the charged decay daughter and the parent direction is evaluated.

The $\tau$ search sensitivity, calculated for 5 years of data taking with a total number of $45 \times 10^{18}$ integrated p.o.t. per year, is given in table~\ref{tab:sensitivity}.

\begin{table}[h]
\vspace{0.4cm}
\begin{center}
\begin{tabular}{|c|c|c|c|}
\hline
$\tau$ decay & \mco{2}{|c|}{Signal $\div \Delta$m$^2$ (Full mixing)}& Background\\
\cline{2-3}
channels & $2.5 \times 10^{-3}$ (eV$^2$)  & $3.0 \times 10^{-3}$ (eV$^2$)& \\ 
 \hline
$\tau \rightarrow \mu^-$& 2.9 &  4.2 & 0.17 \\ 
$\tau \rightarrow e^-$  & 3.5 &  5.0 & 0.17 \\
$\tau \rightarrow h^-$  & 3.1 &  4.4 & 0.24 \\
$\tau \rightarrow 3h $  & 0.9 &  1.3 & 0.17 \\
\hline
ALL                     & 10.4 & 15.0 & 0.76 \\
\hline
\end{tabular}
\caption{\it{Expected number of signal and background events after 5 years of data taking. \label{tab:sensitivity}}}
\end{center}
\end{table}

\begin{center}
\begin{figure}[ht]
  \hspace{3.0cm}
  \includegraphics[scale=0.75]{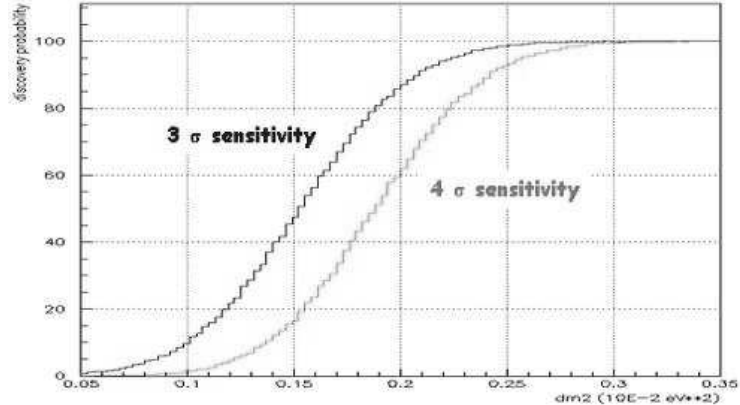}
  \caption{\it 3 and 4 $\sigma$ observation probability as a function of $\Delta$m$^2$.}
    \label{fig:obs_prob}
\end{figure}
\end{center}

\noindent The main background sources are given by:
\begin{itemize}
        \item Large angle scattering of muons produced in $\nu_{\mu}CC$ interactions. 
        \item Secondary hadronic interaction of daughter particles produced at primary $\nu_{\mu}$ interaction vertex. 
        \item Decay of charmed particles produced at primary $\nu_{\mu}$ interaction vertex 
\end{itemize}
Comparing the total number of detected $\nu_{\tau}$ interaction with the estimated background it's clearly seen that OPERA is quite a background-free experiment. In Figure~\ref{fig:obs_prob} the $\nu_{\tau}$ observation probability at 3 and 4 $\sigma$ as a function of $\Delta$m$^2$ is reported.   

\section{Results from the first runs}
The first CNGS run was held in August 2006~\cite{RUN2006}. At that time only electronic detectors were installed: the brick filling started indeed at the beginning of 2007. From 18 to 30 August 2006 a total intensity of $0.76 \times 10^{18}$ p.o.t. was integrated and 319 neutrino-induced events were collected (interactions in the rock surrounding the detector, in the spectrometers and in the target walls). Thanks to this first technical run the detector geometry was fixed and the full reconstruction of electronic detectors data tested. It was also possible
to fine-tune the synchronization between CERN and Gran Sasso, performed using GPS clocks. Furthermore, the zenith-angle distribution from penetrating muon tracks was reconstructed and the measured mean angle of $3.4 \pm 0.3^o$ was well in agreement with the value of $3.3^o$ expected for CNGS neutrinos traveling from CERN to the LNGS underground laboratories.

\begin{center}
\begin{figure}[ht]
  % \vspace{9.0cm}
  \hspace{0.5cm}
  \includegraphics[scale=0.75]{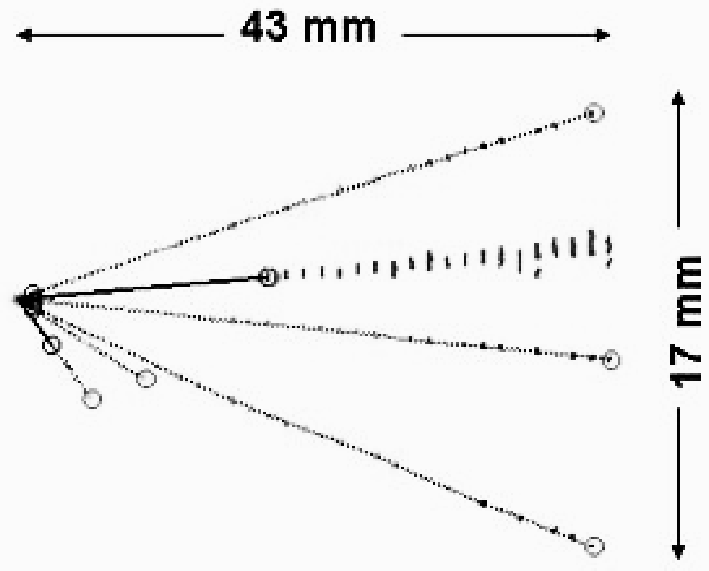}
  \hspace{1cm}
  \includegraphics[scale=0.75]{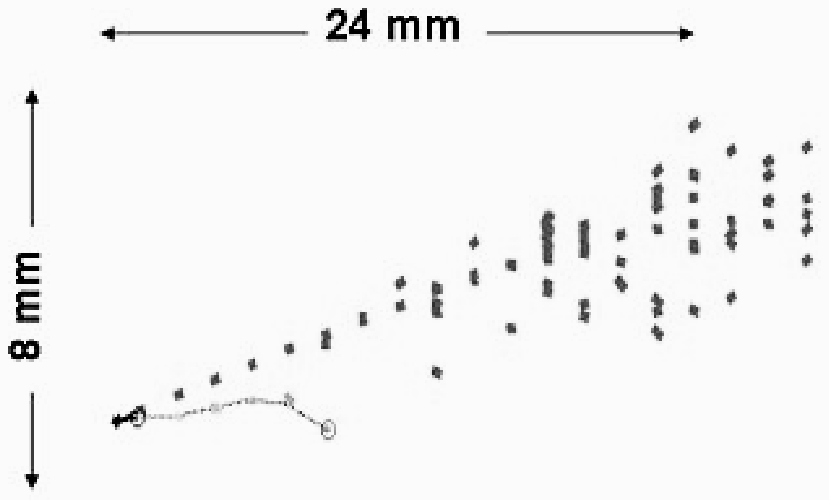}
  \caption{\it Two reconstructed neutrino interaction from the OPERA 2007 run. The event displayed on the left is a a $\nu_{\mu}$CC interaction. The right side shows an event where an energetic shower comes from the interaction vertex.}
    \label{fig:events}
\end{figure}
\end{center}

The first OPERA physical run was held in October 2007. At that time about 40\% of the target was installed, for a total mass of about 550 tons.
In about 4 days of continuous data taking $0.79 \times 10^{18}$  p.o.t. were 
produced at CERN and 38 neutrino interactions in the OPERA target were triggered by the electronic detectors. The corresponding bricks indicated by the brick finding algorithm were extracted and developed after the cosmic ray exposure and their emulsions sent to the scanning laboratories. In a few hours the first neutrino interactions of the OPERA experiment were successfully located and reconstructed. In Figure~\ref{fig:events} the display of two events is shown. The left one is a $\nu_{\mu}$CC interaction with 5 prongs and a shower reconstructed pointing to the primary interaction vertex ($\gamma$ conversion after a $\pi^o$ decay). In the second a quite energetic shower (about 4.7 GeV) coming from the primary interaction vertex is visible.\\

This first physical run was quite short but very significative. Indeed it allowed a full testing of the electronic detectors and the data acquisition. Furthermore, the brick finding algorithm was successfully used to locate the bricks were the neutrino interaction occurred. Finally, the target tracker to brick matching was proved to be able to satisfy the expectations and the full scanning strategy validated.

\section{Outlook and future plans}
The OPERA target will by completed by May 2008. In June a first 150-day period of CNGS beam at nominal intensity is expected to start. About $30 \times 10^{18}$ p.o.t. will be integrated, equivalent to about 3500 neutrino interactions. More then 100 charm decays will be collected, so that the capability to reconstruct $\tau$ decays will be fully exploited. The corresponding number of expected triggered $\nu_{\tau}$ is 1.3: with some luck the first $\nu_{\tau}$ candidate event will be observed during the 2008 OPERA run.     

\end{document}